\documentclass[preprint,12pt]{elsarticle}
\journal{Annals of Physics}

\begin{document}

%\begin{frontmatter}

\title{Possible existence of wormholes in the central regions of halos}

\author{Farook Rahaman}
\address{Department of Mathematics, Jadavpur University, Kolkata
700032, West Bengal, India\\rahaman@iucaa.ernet.in}

\author{P. Salucci}
\address{SISSA,  International School  for  Advanced
Studies, Via Bonomea  265,  34136,  Trieste, Italy  and  INFN,
Sezione  di Trieste,  Via  Valerio  2,  34127, Trieste,
Italy\\salucci@sissa.it}

\author{P.K.F. Kuhfittig}
\address{Department of Mathematics, Milwaukee School of
Engineering, Milwaukee, Wisconsin 53202-3109,
USA\\kuhfitti@msoe.edu}

\author{Saibal Ray}
\address{Department of Physics, Government College of Engineering
\& Ceramic Technology, Kolkata 700010, West Bengal,
India\\saibal@iucaa.ernet.in}

\author{Mosiur Rahaman}
\address{Department of Mathematics, Meghnad Saha Institute of
Technology, Kolkata 700150, India\\mosiurju@gmail.com}

\date{Received: date / Accepted: date}

%\end{frontmatter}

\begin{abstract} An earlier study \cite{fR14,pK14}
has demonstrated the possible existence of wormholes in the outer
regions of the galactic halo, based on the Navarro-Frenk-White
(NFW) density profile. This paper uses the Universal Rotation
Curve (URC) dark matter model to obtain analogous results for the
central parts of the halo. This result is an  important compliment
to the earlier result, thereby confirming the possible existence
of wormholes in most of the spiral galaxies.
\end{abstract}


\maketitle

\section{Introduction}
Wormholes are hypothetical handles or tunnels in spacetime linking
widely separated regions of our Universe or entirely different
universes. Morris and Thorne \cite{MT88} proposed the following
line element for the wormhole spacetime:
\begin{equation}\label{E:line1}
ds^2=-e^{2f(r)}dt^2+\left(1-\frac{b(r)}{r}\right)^{-1}dr^2
+r^2(d\theta^2+\sin^2\theta\,d\phi^2),
\end{equation}
using units in which $c=G=1$.  Here $f=f(r)$ is called the
\emph{redshift function}, which must be everywhere finite to
prevent an event horizon.  The function $b=b(r)$ is called the
\emph{shape function}, which has the property that
$b(r_{th})=r_{th}$, where $r= r_{th}$ is the \emph{throat} of the
wormhole.  A key requirement is the \emph{flare-out condition} at
the throat: $b'(r_{th})<1$, while $b(r)<r$ near the throat. The
flare-out condition can only be satisfied by violating the null
energy condition (NEC), which states that
$T_{\mu\nu}k^{\mu}k^{\nu}\ge 0$ for all null vectors and where
$T_{\mu\nu}$ is the energy-momentum tensor.  So given the null
vector $(1,1,0,0)$, the NEC is violated if $\rho+p_r<0$, where
$\rho$ is the energy density and $p_r$ the radial pressure.

The possible existence of wormholes in the outer region of the
halo has already been discussed in Refs. \cite{fR14,pK14} using
the Navarro-Frenk-White (NFW) density profile \cite{jN96}:
\begin{eqnarray}
 \rho(r)=\frac{\rho_s}{\frac{r}{r_s}\left(1+
    \frac{r}{r_s}\right)^2},\nonumber
\end{eqnarray}
where $r_s$ is the characteristic scale radius and $\rho_s$ is the
corresponding density. This model yields a shape function whose
basic properties, such as the throat size, remain the same in the
region considered \cite{pK14}.  It is well known that the NFW
model predicts velocities in the central parts that are too low
\cite{gG04}, but these discrepancies do not exist in the outer
regions of the halo where the wormholes discussed in Refs.
\cite{fR14,pK14} are located \cite{cT06,aM12}.

In this study we are going to be primarily concerned with the
region closer to the center where the Universal Rotation Curve
(URC) dark matter profile is valid  \cite{salu12}:
\begin{equation}\label{E:rho}
\rho(r)=\frac{\rho_0 r_0^3}{(r+r_0)(r^2+r_0^2)};
\end{equation}
here $r_0$ is the core radius and $\rho_0$ the effective core
density.  While the URC model is valid throughout the halo region,
we assume that the outer region has already been dealt with in
Refs. \cite{fR14,pK14} using the NFW model, thereby leaving only
the central region, which is the subject of this paper.

In this connection we would like to add here that the URC
represents any single rotation curve in spirals of any mass and
Hubble type, and it is an obvious step forward with respect to
assuming a constant value. At some time, a Cored Burkert profile
is a step forward with respect to NFW profile that, it is now
common fact that the latter fails to reproduce the Dark Matter
distribution. Both the URC and the Cored profile are born
empirically and find some explanation later on \cite{S13}.

Therefore, our plan of the present work is as follows: In Sec. 2
we provide the basic equations and their solutions under the URC
dark matter profile whereas Sec. 3 is devoted for some specific
comments regarding the results obtained in the study.

\section{The basic equations and their solutions}
Even though we now have the density profile, other properties of
dark matter remain unknown. So we are going to assume that dark
matter is characterized by the  general anisotropic
energy-momentum tensor \cite{NB}
\begin{equation}
 T_\nu^\mu=(\rho + p_t)u^{\mu}u_{\nu} - p_t g^{\mu}_{\nu}+
            (p_r -p_t )\eta^{\mu}\eta_{\nu},
\end{equation}
with $u^{\mu}u_{\mu} = - \eta^{\mu}\eta_{\mu} = -1$, $p_t$
and $p_r$ being the transverse and radial pressures,
respectively.  The line element for the galactic halo
region is given in Eq. (\ref{E:line1}).

The flat rotation curve for the circular stable geodesic
motion in the equatorial plane yields  the tangential
velocity \cite{SC83,LL75}
 \begin{equation}\label{E:v1}
   (v^{\phi})^2= r  f'(r).
\end{equation}

The radius $r$ in kpc and velocity $v^\phi$ in km/s of the
Rotation Curve of objects with total virial mass $3\times10^{12}$
solar masses is given below (Table - 1) \cite{S13}. We find the
best fitting curve which is given in Fig. 1.

By applying intuition, we propose that the observed rotation curve
profile in the dark matter region is of the form
 \begin{equation}\label{E:v2}
   v^{\phi} = \alpha r \exp (- k_1 r)
   +\beta[1 - \exp (- k_2 r)].
\end{equation}
For a typical galaxy the tangential velocity $v^\phi$ is shown in
Fig. 1.  Moreover, for sufficiently large $r$, $v^{\phi}\sim $~
250 ~ km/s $\approx 0.00083 $ \cite {FR08,FR09}. One can note that
our proposed curve and observed curve profile for tangential
velocity are almost similar to each other for the specific values
of the parameters i.e. the proposed and observed rotational
velocities are both fittable with our empirical formula.
Therefore, our assumption is more or less justified.

\begin{figure*}[thbp]
\begin{tabular}{rl}
\includegraphics[width=5.5cm]{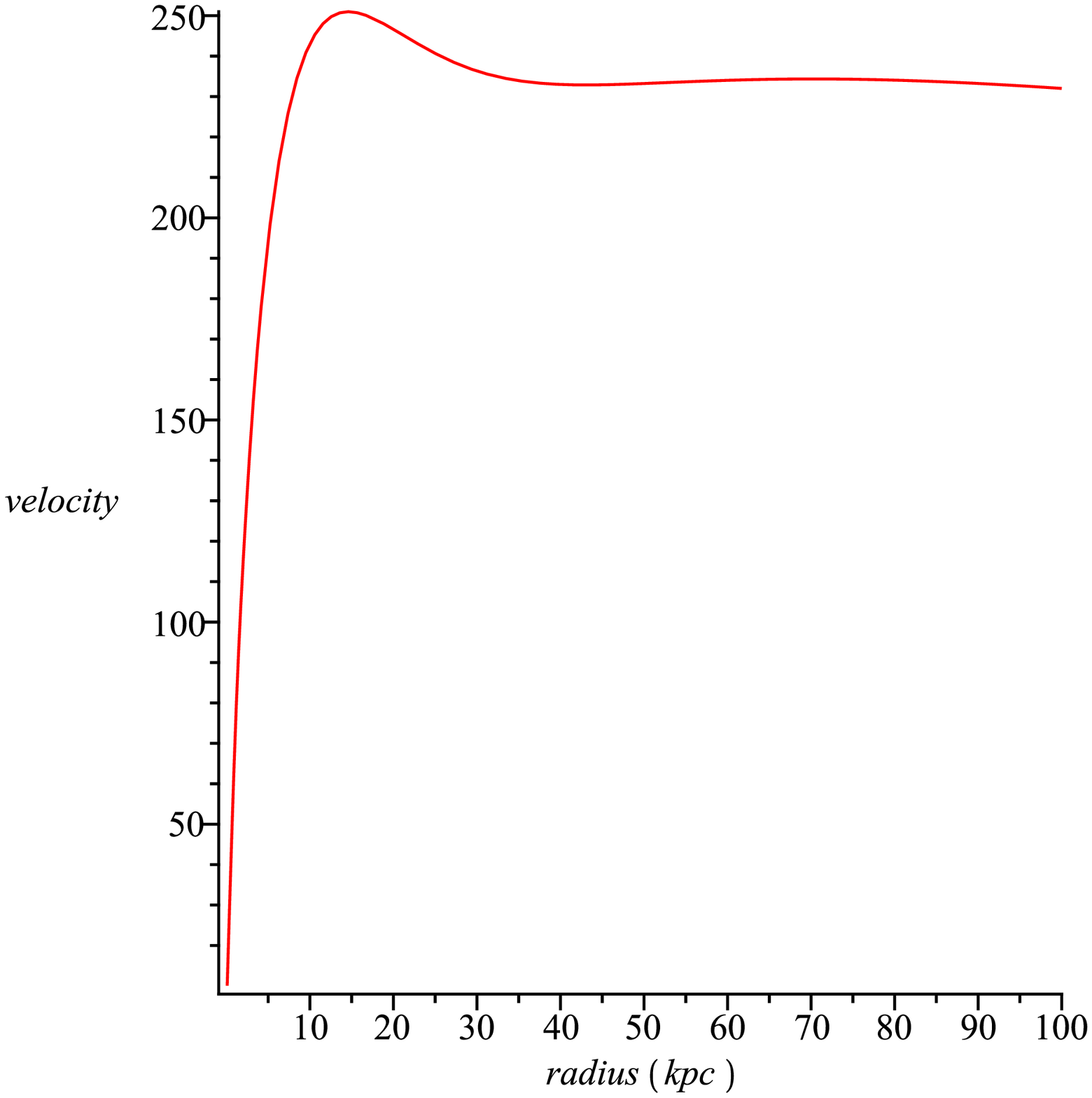}&
\includegraphics[width=5.5cm]{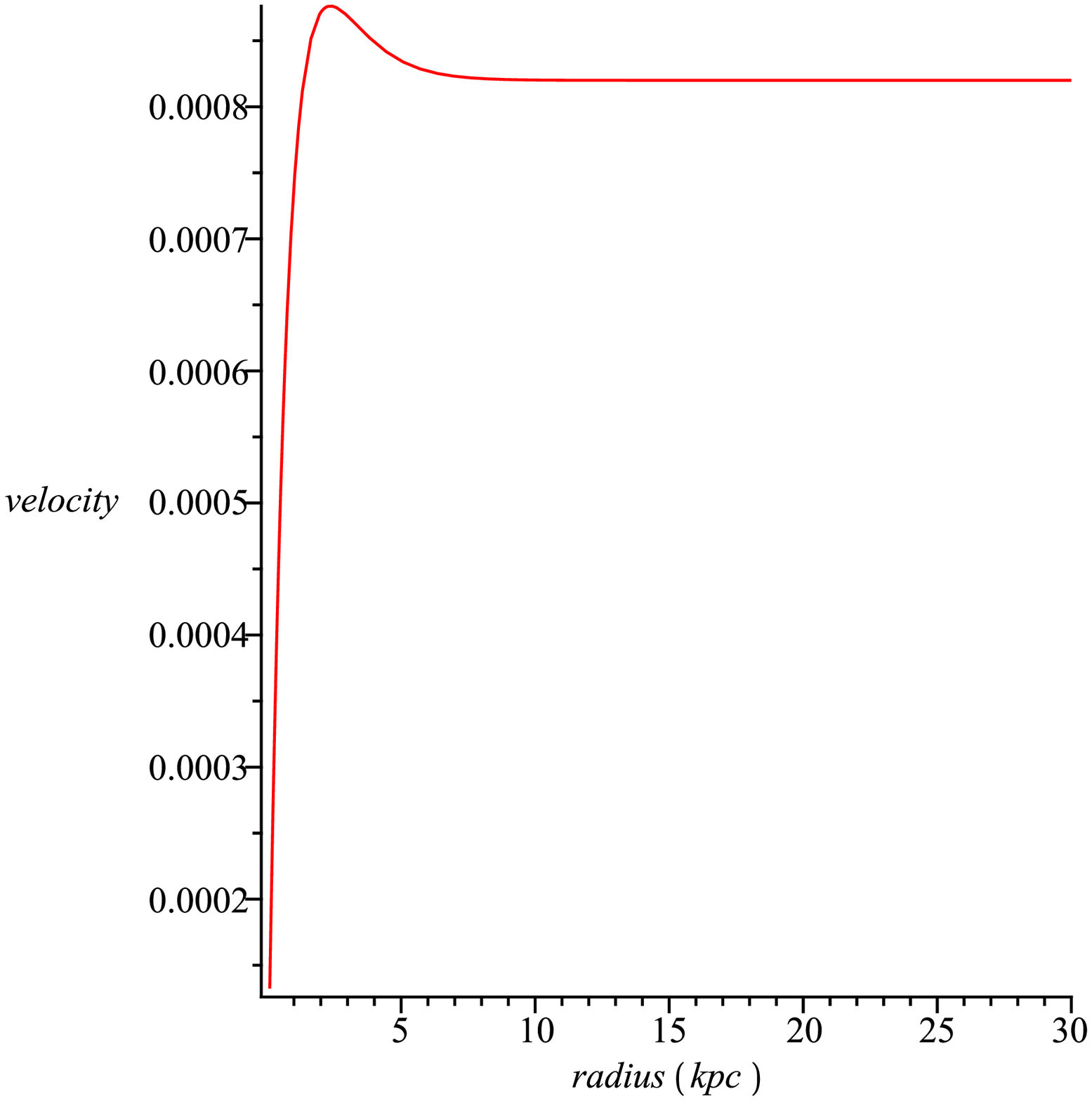} \\
\end{tabular}
\caption{ (Left) The rotational velocity from observations.
(Right) The proposed rotational velocity with the values of the
parameters as $k_1=k_2=1, \alpha=0.0006, \beta=0.00082$. }
\end{figure*}

\begin{table}
\caption{The radius $r$ in kpc and velocity $v^\phi$ in km/s of
the Rotation Curve of objects with total virial mass $3\times
10^{12}$  solar masses.} \label{tab3} \centering
\bigskip
{\small

\begin{tabular}{llllll}
\hline \\[-9pt]

$R$ (kpc) & $V$ (km/s) &~ $R$ (kpc) & $V$ (km/s)  \\

\hline \\[-9pt] 0.1 & 10.052834744388612 & 26.1 &
239.5389556726835\\

\hline \\[-9pt] 1.1 & 74.46665498068327 & 27.1 &
238.58964093528954\\

\hline \\[-9pt] 2.1 & 118.22330294498593 & 28.1 &
237.72382676461396 \\

\hline \\[-9pt] 3.1 & 151.1128881336094 & 29.1 &
236.94252666484462\\

\hline \\[-9pt] 4.1 & 176.44533836937705 & 30.1 &
236.24478171375338 \\

\hline \\[-9pt] 5.1 & 196.0988184186591 & 31.1 &
235.62813232944393 \\

\hline \\[-9pt] 6.1 & 211.33095938495612 & 32.1 &
235.08901162538888  \\

\hline \\[-9pt] 7.1 & 223.05674521744098 & 33.1 &
234.62307027076736\\

\hline \\[-9pt] 8.1 & 231.97462265975628 & 34.1 &
234.62307027076736\\

\hline \\[-9pt] 9.1 & 238.63397717705232 & 35.1 &
234.22544187966773\\

\hline \\[-9pt] 10.1 & 243.47545965729188 & 36.1 &
233.89095712722872\\

\hline \\[-9pt] 11.1 & 246.8570877004846 & 37.1 &
233.39021089298888\\

\hline \\[-9pt] 12.1 & 249.07218340068806 & 38.1 &
233.2134483391877\\

\hline \\[-9pt] 13.1 & 250.36228653986132 & 39.1 &
233.07900487106537\\

\hline \\[-9pt] 14.1 & 250.92679861285606 & 40.1 &
232.98209135254623\\

\hline \\[-9pt] 15.1 & 250.93040201908613 & 41.1 &
232.91818791468324 \\

\hline \\[-9pt] 16.1 & 250.5089080065525 & 42.1 &
232.88306686283582\\

\hline \\[-9pt] 17.1 & 249.77395990965775 & 43.1 &
232.87280446959267 \\

\hline \\[-9pt] 18.1 & 248.81687987924815 & 44.1 &
232.88378410156642\\

\hline \\[-9pt] 19.1 & 247.71185949364315 & 45.1 &
232.91269272147773 \\

\hline \\[-9pt] 20.1 & 246.51863692196736 & 46.1 &
232.95651245084068\\

\hline \\[-9pt] 21.1 & 245.2847642472647 & 47.1 &
233.0125085700085\\

\hline \\[-9pt] 22.1 & 244.04754149102484 & 48.1 &
233.07821506796643 \\

\hline \\[-9pt] 23.1 & 242.8356747965165 & 49.1 &
233.15141862990475\\

\hline \\[-9pt] 24.1 & 241.67070261276308 & 50.1 &
233.23014176204364 \\

\hline \\[-9pt] 25.1 & 240.56822394226268 & \\

\hline
\end{tabular}   }
\end{table}

\pagebreak

The Einstein field equations for the above metric are
\begin{eqnarray}
\frac{b^{\prime }(r)}{ r^{2}}\label{E:Ein1}&=&8\pi\rho (r),\\
 2\left(1-\frac{b}{r}\right) \frac{f^\prime}{r}
 -\frac{b}{r^{3}}\label{E:Ein2}&=&8\pi
p_r(r),\\ \left(1-\frac{b}{r}\right)\left[ f^{\prime \prime}+
\frac{f^\prime}{r} +{f^\prime}^2 - \left\{
\frac{b^{\prime }r-b}{2r(r-b)}\right\}\left(f^{\prime} +
\frac{1}{r}\right)\right]\label{E:Ein3}&=&8\pi p_{t}(r).
\end{eqnarray}
From Eqs. (\ref{E:v1}) and (\ref{E:v2}) and using some
typical values, we obtain the redshift function

\begin{eqnarray}\label{E:redshift}
f(r)  =  -\frac{\alpha^2 r}{2 k_1 e^{(2 k_1 r)}}
-\frac{\alpha^2}{4k_1^2 e^{(2k_1r)}} -\frac{2\alpha\beta}{k_1
e^{(k_1r)}} + \frac{2\alpha\beta e^{(-k_1r-k_2r)}}{k_1 + k_2}
\nonumber \\ +\beta^2\ln(r)+2\beta^2 E_i(1,k_2r)-\beta^2
E_i(1,2k_2 r)+D .
\end{eqnarray}
Here $E_i(*,*)$ is the exponential integral and $D$ is an
integration constant.  The graph of $f(r)$ in Fig. 2 shows the
behavior in the central part of the halo, which is the region that
we are primarily concerned with. (For large $r$, $f(r)$ is such
that $e^{2f(r)}=B_0r^l$ where $l=2(v^{\phi})^2$ \cite{FR08,FR09}).

The determination of the shape function $b=b(r)$ requires
a little more care.  First of all, we assume that $b(r)$,
$f(r)$, and the halo have a common origin in order to use
Eqs. (\ref{E:v1}) and (\ref{E:Ein1}) in the calculations.
To see if the shape function meets the basic requirements,
we start with Eq. (\ref{E:Ein1}),
\[
    b'(r)=8\pi r^2\rho(r),
\]
and integrate from 0 to $r$ to obtain
\begin{equation}
 b(r)= 4\pi r_0^3\rho_0\ln(r+r_0)+2\pi r_0^3 \rho_0\ln(r^2+r_0^2)
 -4\pi\rho_0r_0^3\tan^{-1}\left(\frac{r}{r_0}\right)+C,
\end{equation}
where $C$ is an integration constant.  To get an overview of the
shape function, we assign some arbitrary values to the parameters
and obtain the plots in Fig.2  shows that the throat is located
at some $r=r_{th}$, where $b(r)-r$ intersects the $r$-axis.
Also, for $r>r_{th}$, $b(r)-r<0$, which implies that $b(r)/r<1$.
Furthermore, $b(r)-r$ is decreasing for $r\ge r_{th}$, implying
that $b'(r_{th})<1$.  Hence the flare-out condition is
satisfied.  So based on the URC model, the qualitative features
meet all the requirements for the existence of a wormhole.

At this point it would be desirable to examine the effect
of using more specific parameters.  For example, if the
throat of the wormhole coincides with the core radius
$r=r_0$, then we get for the shape function
\begin{eqnarray}
 b(r)  = \int_{r_0}^r 8 \pi r^2 \rho dr +r_0 ~~~~~~~~~~~~~~~~~~~~~~~~~~~~~~~~~~~~~~~\nonumber \\
=4\pi r_0^3\rho_0\ln(r+r_0)+2\pi r_0^3 \rho_0\ln(r^2+r_0^2)-4\pi
\rho_0r_0^3\tan^{-1}\left(\frac{r}{r_0}\right)+C,
\end{eqnarray}
where
\begin{eqnarray}C = r_0 -4\pi
r_0^3\rho_0 \ln 2r_0 - 2\pi r_0^3\rho_0 \ln 2r_0^2 + \pi^2
\rho_0r_0^3. \nonumber \end{eqnarray}
As before, $b(r_0)=r_0$.

Having determined the throat radius allows a closer look at
the flare-out condition, which again depends on $b'(r)$.
By Eqs. (\ref{E:Ein1}) and (\ref{E:rho}),
\begin{equation}\label{E:bprime}
   b'(r)=8\pi r^2\frac{\rho_0r_0^3}{(r+r_0)(r^2+r_0^2)}.
\end{equation}
In the specific case of the Milky Way galaxy, $r_0=9.11$~kpc
\cite{aM12} and $\rho_0=5\times 10^{-24}(r_0/8.6 $~kpc)$^{-1}$
g~cm $^{-3}$ \cite{salu12}. Substituting in Eq. (\ref{E:bprime}),
we get
\[
   b'(r_0)\approx 1.74\times 10^{-6}.
\]
The flare-out condition is easily satisfied . Notice that this
result does not change if we adopt the   Milky Way values for
$rho_0$ and $r_0$  found by the accurate analysis of \cite{NS}

To complete our discussion, we will use Eqs. (\ref{E:Ein2})
and (\ref{E:Ein3}) to obtain the radial and lateral pressures:

\begin{eqnarray} p_r(r)  =  \frac{1}{8\pi r^3}\left[2 (r - A)B^2 -
A\right],
 \end{eqnarray}

$ p_t(r)  = \frac{1}{8\pi}\left(1-\frac{A}{r}\right) \times
\nonumber \\ \left[\frac{2B \{\alpha e^{(-k_1r)}-\alpha rk_1
e^{(-k_1r)}+\beta k_2 e^{(-k_2 r)}\}}{r} + \frac{B^4}{r^2} -
\frac{\left(\frac{8\pi\rho_0 r^6}{(r+r_0)(r^2+r_0^2)}-
A\right)(B^2+\frac{1}{r})}{2r^2(r - A)} \right] $.

\begin{eqnarray} \end{eqnarray}

where
\begin{eqnarray}
A = 4\pi r_0^3\rho_0\ln(r+r_0)+2\pi r_0^3
\rho_0\ln(r^2+r_0^2)-4\pi
r_0^3\rho_0\tan^{-1}\left(\frac{r}{r_0}\right)+C \nonumber,
\end{eqnarray}

\begin{eqnarray}
B =  v^{\phi} = \alpha r e^{(-k_1r)} +
\beta[1-e^{(-k_2r)}]\nonumber.
\end{eqnarray}

According to Fig. 3, $\rho+p_r<0$, so that the null energy
condition is indeed violated.

\begin{figure}[htbp]
    \centering
        \includegraphics[scale=.3]{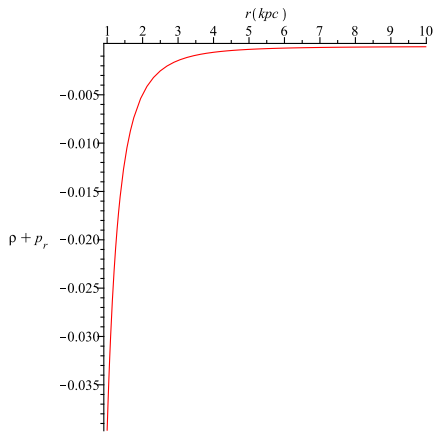}
       \caption{Plot showing that $\rho+p_r<0$ for values of the parameters $r_0=1, \rho_0=0.0001$}
    \label{fig:8}
\end{figure}

\section{Concluding remarks}
The possible existence of wormholes in the outer regions of the
halo was discussed in Ref. \cite{fR14}, based on the NFW density
profile.  Possible detection by means of gravitational lensing,
discussed in Ref. \cite{pK14}, has shown that the basic features,
such as the throat size, are independent of the position within
the halo. Because of certain discrepancies discovered near the
center, this paper uses the observationally motivated URC dark
matter density profile instead in order to accommodate the region
closer to the center of the halo. This center is also the center
of the wormhole.

It is subsequently shown that the solution obtained satisfies all
the criteria for the existence of a wormhole, particularly the
violation of the null energy condition. In the special case of the
Milky Way galaxy, if the throat radius $r=r_0$ coincides with the
radius of the core, then $b'(r_0)\approx 1.74\times 10^{-6}$,
thereby meeting the flare-out condition.  A similar result can be
expected for most spiral galaxies. Thus our result is very
important because it confirms the possible existence of
 wormholes in  most of the  spiral galaxies. Scientists remain silent on whether
it is possible to manufacture or create of the exotic matter
violating null energy condition  in laboratory. As a result the
construction of a  wormhole geometry in our real world is
extremely difficult. However in the  galactic halo region, dark
matter may supply the fuel for constructing and sustaining  a
wormhole. Hence, wormholes could be found in nature and our study
may encourage scientists to seek observational evidence for
wormholes in the galactic halo region.

\section*{Acknowledgments} FR and SR are thankful to the
Inter-University Centre for Astronomy and Astrophysics (IUCAA),
India for providing Associateship Programme where a part of the
work has been done. FR is also grateful to Jadavpur University,
India for financial support.

\end{document}